\newcommand{\figta}{$\left(\mathrm{a}\right)\;$}
\newcommand{\figtb}{$\left(\mathrm{b}\right)\;$}
\newcommand{\figtc}{$\left(\mathrm{c}\right)\;$}

\newcommand{\figa}{$\left(\mathrm{a}\right)$}
\newcommand{\figb}{$\left(\mathrm{b}\right)$}
\newcommand{\figc}{$\left(\mathrm{c}\right)$}

\newcommand{\equref}[1]{(\ref{gap})}

\newcommand{\mr}[1]{\mathrm{#1}}
\newcommand{\be}{\begin{equation}}
\newcommand{\ee}{\end{equation}}

\RequirePackage{fix-cm}
\documentclass[smallextended]{svjour3}       

\smartqed  

\usepackage{graphicx}

\begin{document}

\title{Magnetometry with low resistance proximity Josephson junction} 

\author{R. N. Jabdaraghi, J. T. Peltonen,  D. S. Golubev,  and  J. P. Pekola} 

\institute{R. N. Jabdaraghi, J. T. Peltonen,  D. S. Golubev,  and  J. P. Pekola\\
Low Temperature Laboratory, Department of  Applied  Physics,\\
 Aalto University, FI-00076 AALTO, FINLAND\\
E-mail: robab.najafi.jabdaraghi@aalto.fi }









\maketitle
\begin{abstract}
We characterize a niobium-based superconducting quantum interference proximity transistor (Nb-SQUIPT) built upon a Nb-Cu-Nb SNS weak link. The Nb-SQUIPT and SNS devices are fabricated simultaneously  in two separate lithography and deposition steps, relying on Ar ion cleaning of the Nb contact surfaces. The quality of the Nb-Cu interface is characterized by measuring the temperature-dependent equilibrium critical supercurrent of the SNS junction. In the Nb-SQUIPT device, we observe a maximum flux-to-current transfer function value of about $ 55~\mr{nA/\Phi_0}$ in the sub-gap regime of bias voltages. This results in  suppression of power dissipation down to a few $\mr{fW}$. The device can implement  a low-dissipation SQUIPT,  improving by up to two orders of magnitude compared to a conventional  device based on an Al-Cu-Al SNS junction and an Al tunnel probe (Al-SQUIPT).

\keywords{Proximity effect, SQUIPT, SNS junction,  Nb-SQUIPT}
\end{abstract}

\newpage
\section{Introduction}

The superconducting quantum interference proximity transistor (SQUIPT)~\cite{Giazotto2010} is a sensitive magnetometer based on the proximity effect~\cite{Gennes1964,Gennes1966,Belzig1999,Lambert1998,Pannetier2000,Buzdin2005}. The SQUIPT consists of a superconducting loop ($\mr {S_1}$) interrupted by a normal metal (N) in clean contact with it while a superconducting probe ($\mr {S_2}$) is tunnel coupled to the middle of the weak link. 
\begin{figure}
  \includegraphics[width=1\textwidth]{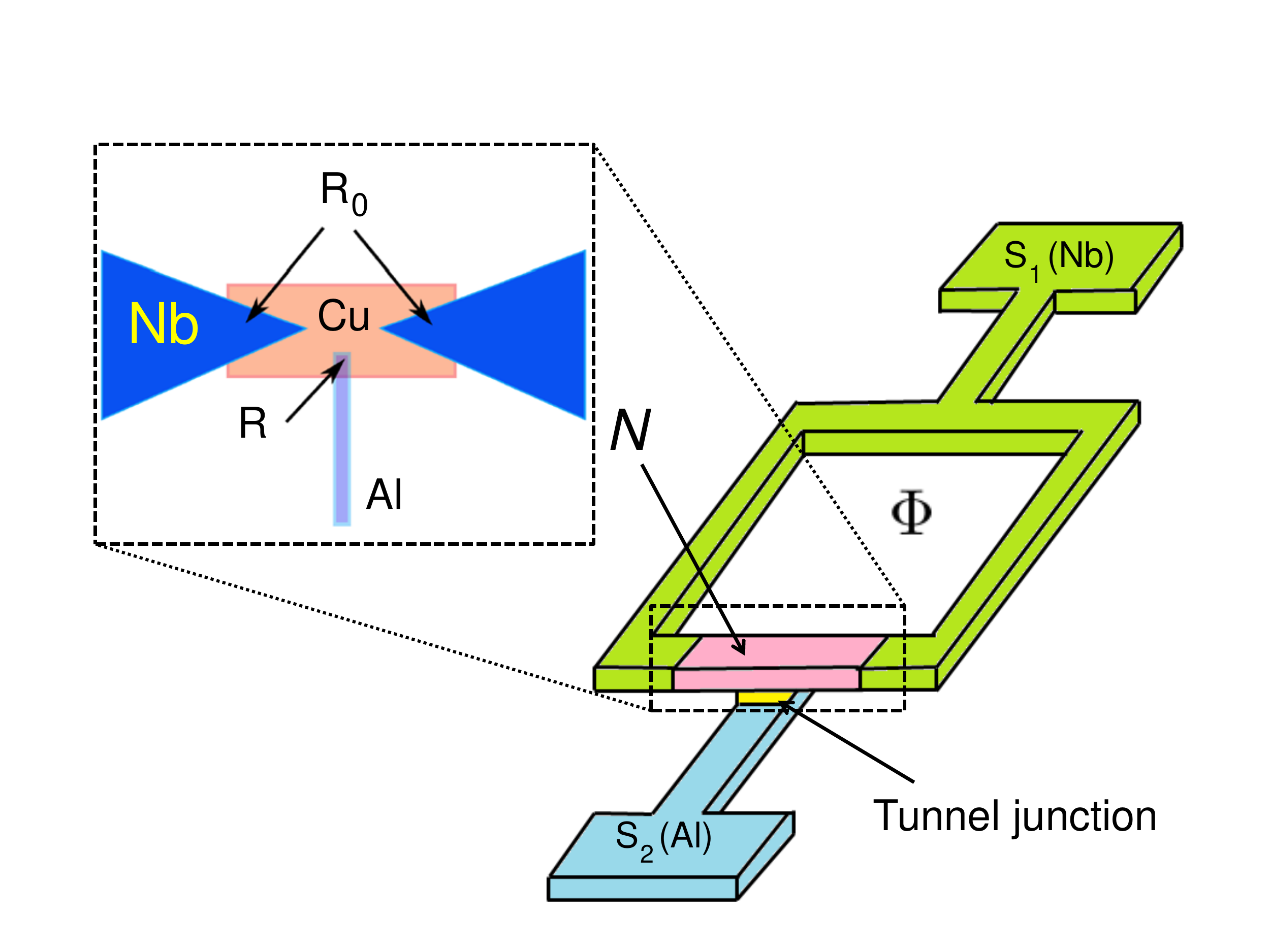}
\centering
\caption{Schematic view of a Nb-SQUIPT device, consisting of a normal metal Cu embedded in a superconducting Nb loop while a superconducting Al probe is tunnel coupled to the middle of the Cu normal metal. The resistance of the contact between Nb and Cu is $R_0$, whereas $R$ denotes the resistance of the junction between Al and Cu. } 
\label{fig:theory}       
\end{figure}
The operation of the device relies on the magnetic field modulation of the DoS \cite{Belzig1999,Belzig1996,Gueron1996,Sueur2008} in the proximized normal metal (N)~\cite{Petrashov1994,Petrashov1995,Belzig2002} embedded between superconducting electrodes ($\mr {S_1}$), resulting in the opening of a minigap ~\cite{Belzig1999,Belzig1996,Zhou1998} in the N part. As shown in Fig.~\ref{fig:theory}, the superconducting loop and probe are Nb and Al, respectively.  
 A large number of  applications  has been proposed for SQUIPTs, similar to SQUIDs~\cite{Clarke2004,Tinkham1996,Likharev1986}, including measurement of magnetic flux induced by atomic spins, single-photon detection and nanoelectronical measurements~\cite{Foley2009,Hao2005,Hao2007}. 
The SQUIPT represents a sensitive interferometer with reduced power dissipation and potentially low flux noise in contrast to conventional SQUIDs~\cite{Giazotto2010,Giazotto2011}. 
So far, a number of SQUIPT devices with an aluminum tunnel probe (S-SQUIPT)~\cite{Giazotto2010,Meschke2011,Najafi2014,Ronzani2014,Najafi2016,Najafi2017,Ligato2017,Ronzani2017,Ambrosio2015} have been reported due to the high quality native oxide of Al, yielding excellent tunnel barriers~\cite{Ligato2017}. 

The flux sensitivity of a SQUIPT device can be improved by maximizing the proximity effect in the normal metal N, hence increasing the responsivity of the device  to a change in magnetic field. In this respect, replacing the superconducting loop with a larger-gap superconductor such as vanadium~\cite{Ligato2017} or niobium~\cite{Najafi2016} is an evident direction to look for sensitive SQUIPTs. Recently,  a device with normal-conducting probe (N-SQUIPT) has been shown to be a promising  candidate for the implementation of SQUIPTs with very low dissipated power~\cite{Ambrosio2015}. However, implementing low dissipation S-SQUIPTs has not been explored in detail up to now. 
Operating the S-SQUIPT at sub-gap bias voltages on the supercurrent branch  allows to obtain extra-low power dissipation, reduced by up to two orders of magnitude compared to earlier devices~\cite{Giazotto2010,Najafi2014}. In this article, we present a detailed experimental demonstration of  Nb-Cu-Nb weak links and a SQUIPT based on a Nb superconducting loop (Nb-SQUIPT), first considered in Ref. 25. The devices are realized using  an etching-based two-step fabrication process. We first characterize Nb-Cu-Nb junctions through low temperature switching current measurement and then investigate a Nb-SQUIPT, obtained by attaching an Al tunnel probe to the SNS weak link.   
 In this prototype magnetometer, we observe the maximum flux-to-current transfer function value of about $|\partial I/\partial \Phi|_\mr{max}\approx 55~ \mr{nA/\mr{\Phi_0}}$ at $T=80~\mr{mK}$, which has been measured close to zero bias voltage. Here, $\mr{\Phi_0}=h/(2e)$ is the superconducting flux quantum.

\section{Experimental methods and fabrication}
\begin{figure}
  \includegraphics[width=1\textwidth]{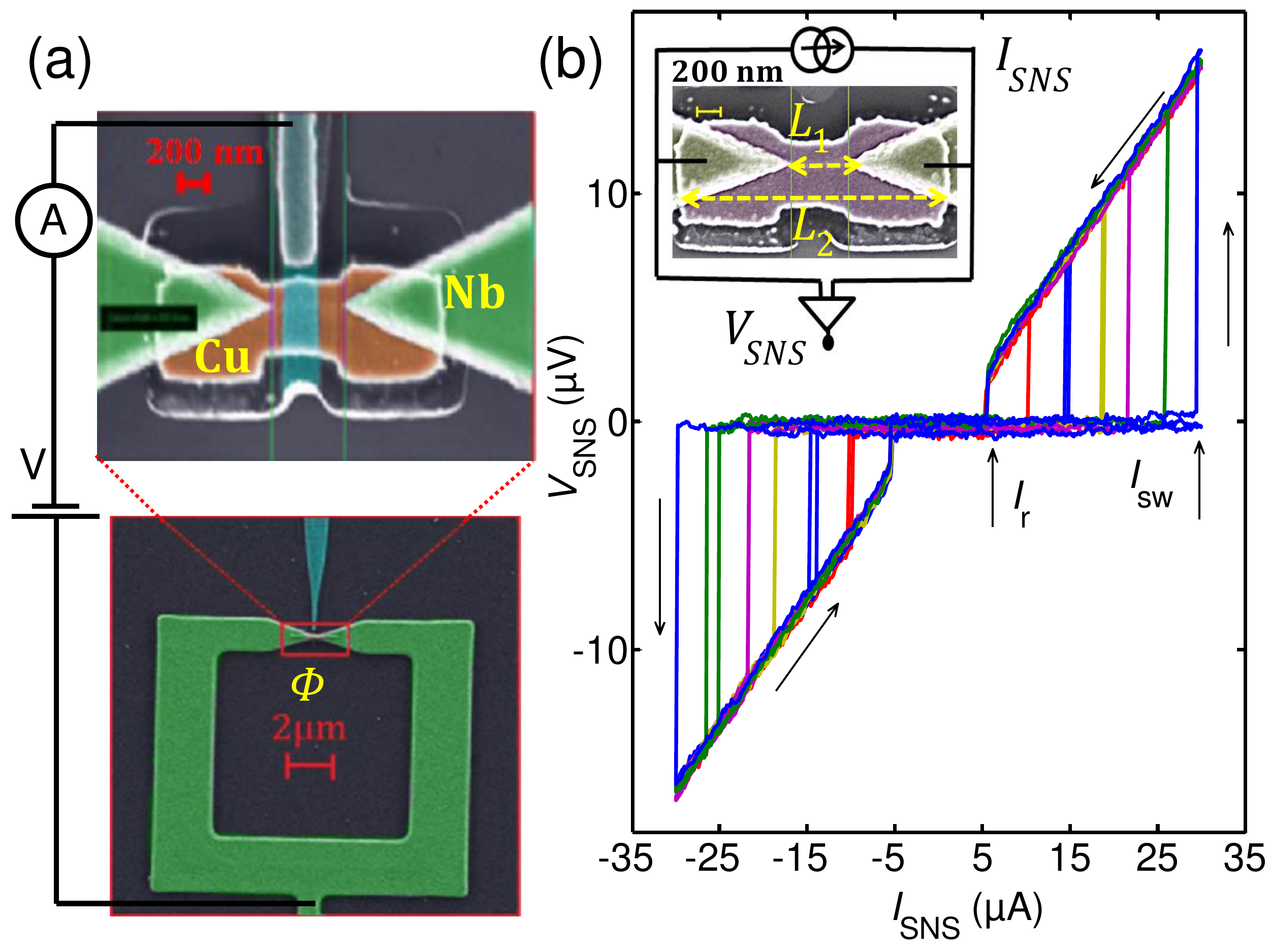}
\centering
\caption{False color scanning electron  micrograph of a Nb-SQUIPT showing (a) an enlarged view of the NIS junction region,  together with the  measurement scheme under voltage bias. (b) Current-voltage (IV) characteristics of a  SNS structure measured at different bath temperatures below 500 mK. Arrows indicate the corresponding switching $I_\mr{sw}$ and retrapping $I_\mr{r}$ currents. The inset shows a representative SEM image of a SNS weak link, consisting of a normal metal (Cu) wire embedded between two superconducting electrodes (Nb). $L_1$ is the distance between the superconducting electrodes, whereas $L_2$ is the total normal metal length.} 
\label{fig:SEM}       
\end{figure}

\begin{figure}
  \includegraphics[width=1\textwidth]{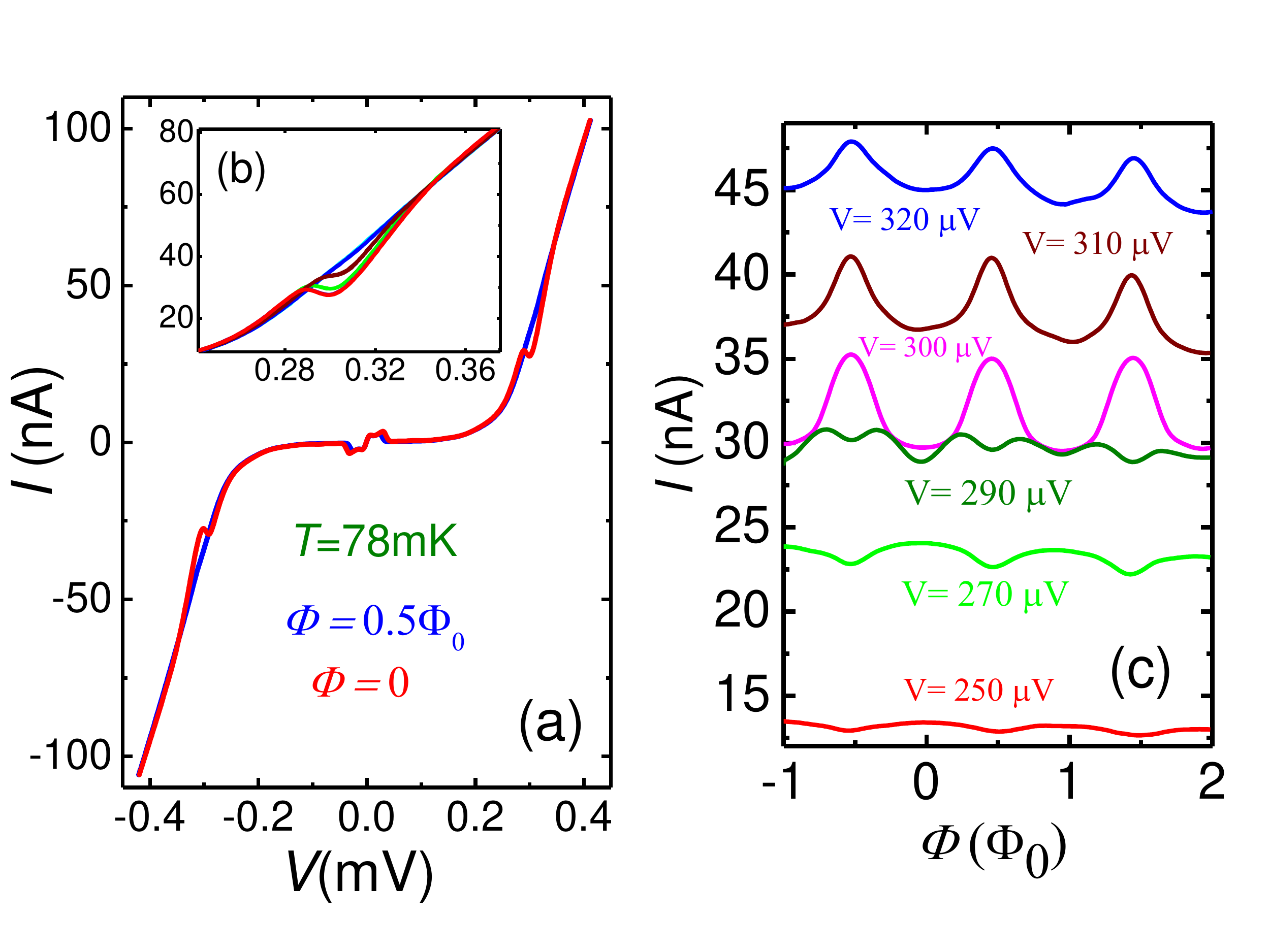}
\centering
\caption{Current-voltage  characteristics of a Nb-SQUIPT device measured at $T=78$ mK for two values of the external magnetic flux through the superconducting loop, $\mr\Phi=0$ (red) and $\mr\Phi=0.5 ~\mr\Phi_0$ (blue). The inset illustrates the extent of the flux modulation of the IV curve around the onset of the quasiparticle current at four selected magnetic flux values in the range  $0<\mr\Phi<0.5~\mr\Phi_0$. (c) Current modulation $I (\mr\Phi)$ for several values of bias voltage in the  bias regime between 250 and 320 $\mu$V.} 
\label{fig:IV}      
\end{figure}

\begin{figure}
  \includegraphics[width=1\textwidth]{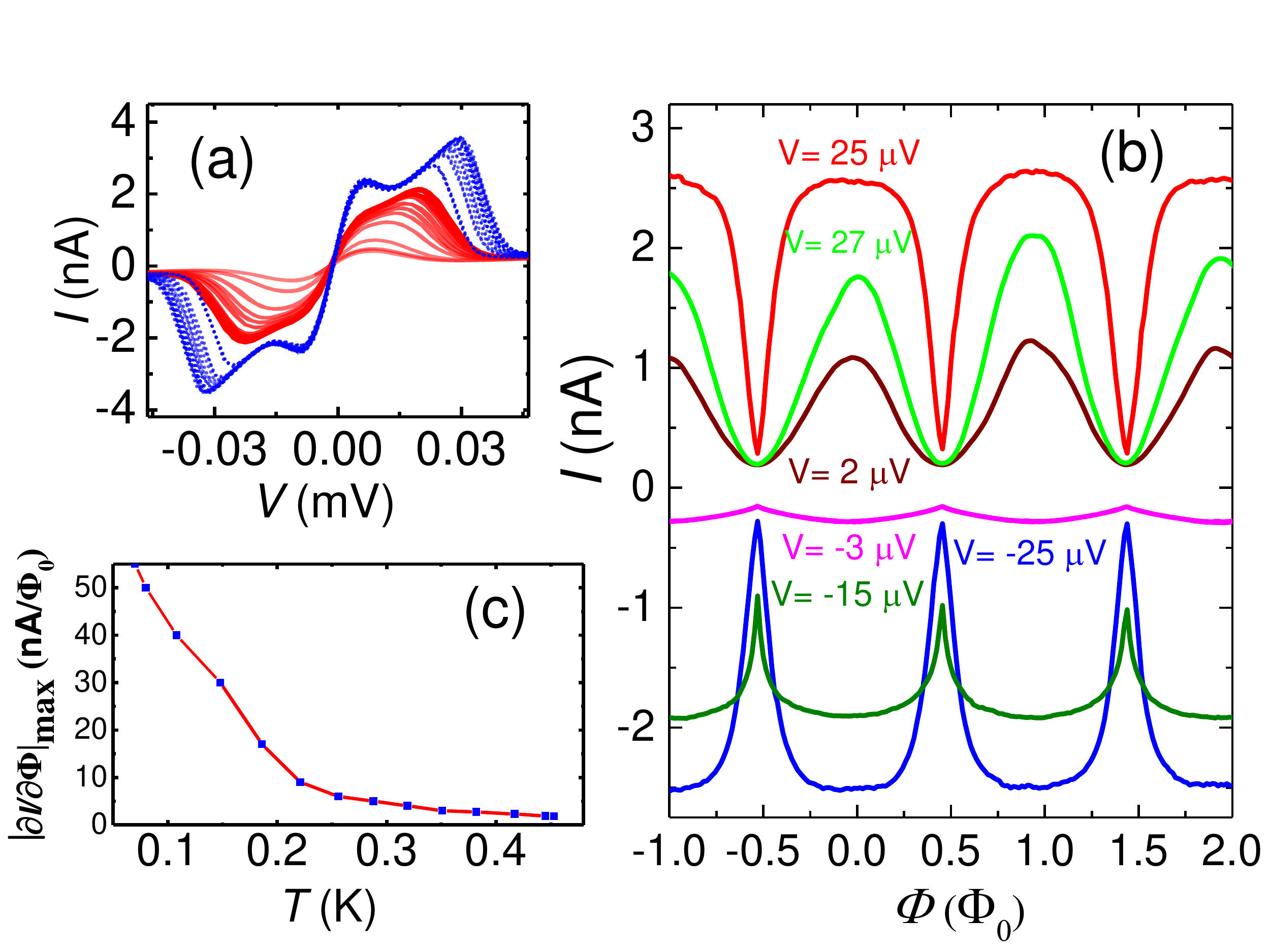}
\centering
\caption{Flux modulation of the IV curve in the sub-gap region at two bath temperatures $T=80$ mK (blue) and $T=190$ mK (red). (b) $I (\mr\Phi)$ at several bias voltages  between -25 $\mu$V (blue) and 25  $\mu$V (red) at $T=190$ mK. (c) Temperature dependence of the maximum flux-to-current transfer function $|\partial I/\partial\Phi|_\mr{max}$ in the sub-gap region of bias voltages.} 
\label{fig:subgap}       
\end{figure}
\begin{figure}
  \includegraphics[width=1\textwidth]{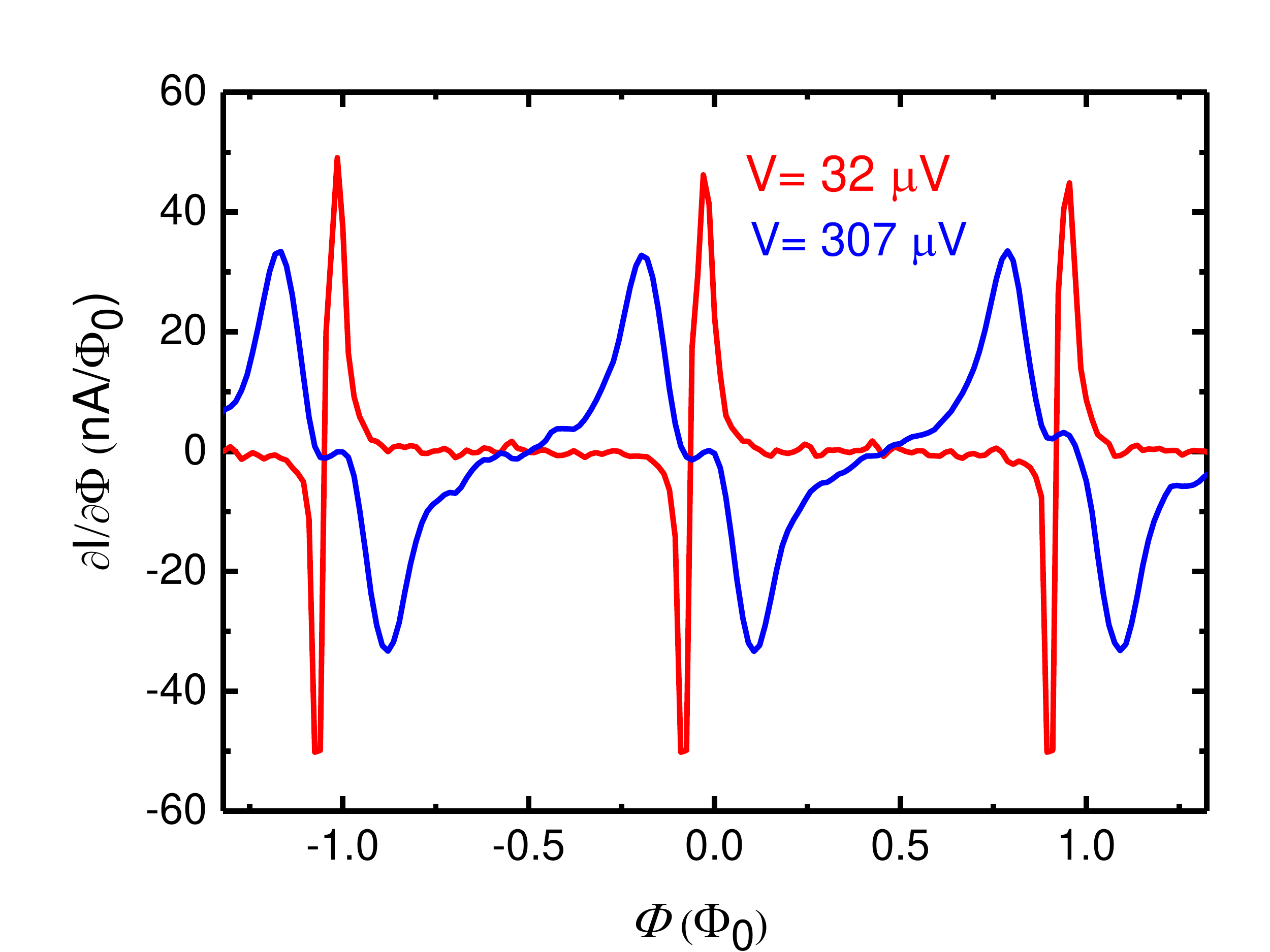}
\centering
\caption{Flux-to-current transfer function  measured at two different bias voltages, in the supercurrent regime at  $V=32~\mr {\mu V}$ and in the quasiparticle branch at $V=307~\mr {\mu V}$.} 
\label{fig:res_flux}       
\end{figure}
\begin{figure}
  \includegraphics[width=1\textwidth]{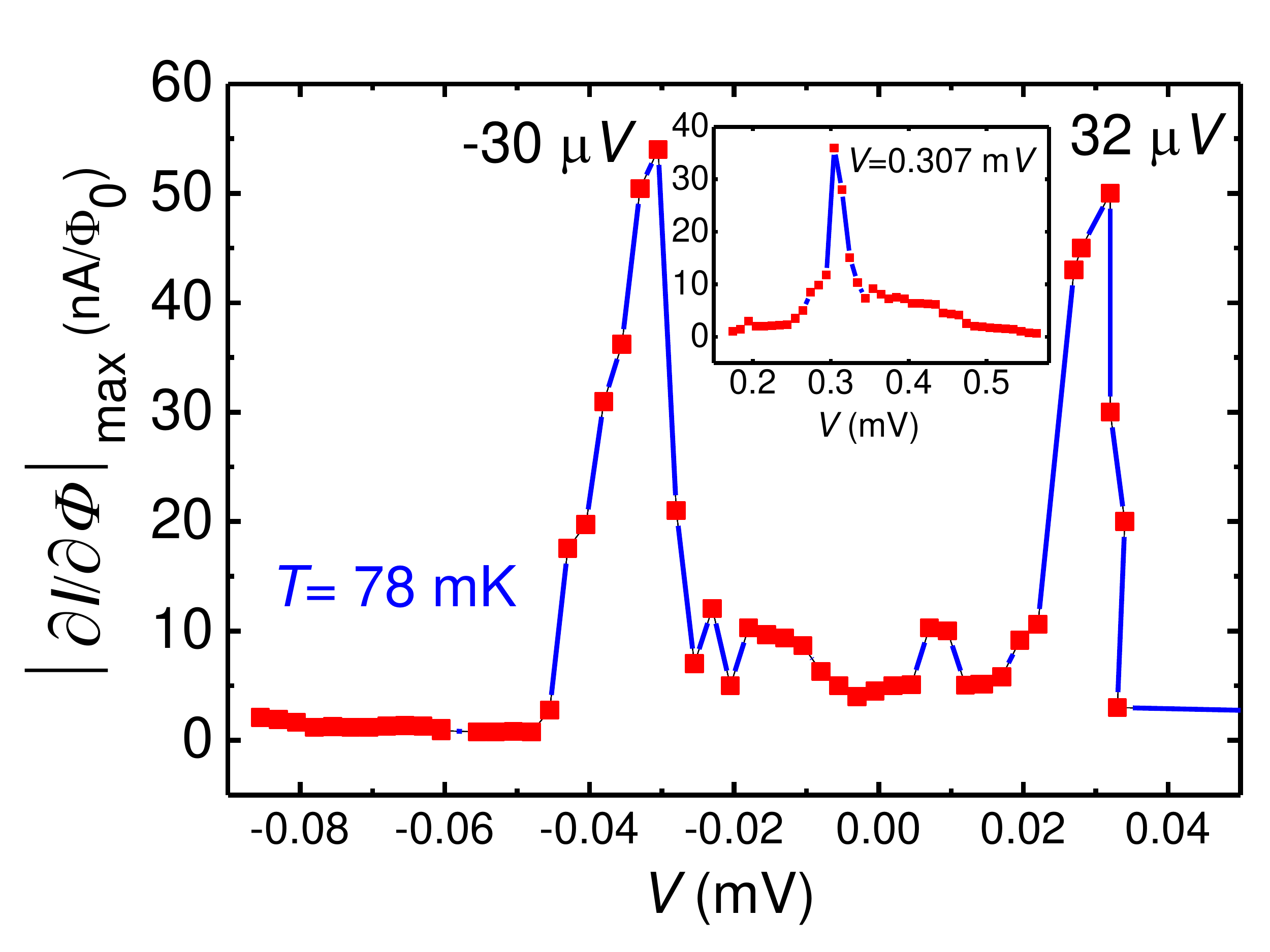}
\centering
\caption{Maximum flux-to-current transfer function as a function of bias voltage,  measured at $T=78$ mK.} 
\label{fig:res_vbias}       
\end{figure}

 Both  Nb-SQUIPT devices and  Nb-Cu-Nb junctions on  the same chip are fabricated using a process with two separate electron beam lithography (EBL) steps as discussed in detail in Ref.~25.  The first step accomplishes patterning of the Nb structures. We start with an oxidized 4 inch Si wafer with 200 nm  sputter-deposited Nb.  In the first lithography step, we write into a layer of spin-coated  positive tone AR-P 6200.13 resist. The lithography is followed by wet development and 5 min reflow baking at  $150 ^\circ \mr C $. In order to transfer the exposed pattern into the Nb film, reactive ion etching (RIE)  is used with a mixture of $\mr{SF_6}$ and Ar. Nb etching is followed by the second EBL step  using a conventional bilayer resist~\cite{Dolan1977},  consisting of a 200 nm thick polymethyl methacrylate (PMMA) layer on top of a 900 nm layer of copolymer. This step provides the NIS tunnel junction and the Cu wire, and  importantly, the clean electrical contacts between Nb and Cu in the Nb-Cu-Nb  weak link. 

The creation of a transparent contact between Nb and Cu is achieved by exposing the chip to $in~situ$ Ar ion etching, prior to the Cu deposition in the same vacuum cycle.  The Nb-Cu-Nb weak link and Nb-SQUIPT devices with an Al-AlOx-Cu Normal metal$-$Insulator$-$Superconductor (NIS) tunnel junction are simultaneously fabricated: After Ar etching, 25 nm of Al at an angle $\theta=27^\circ$ is deposited and oxidized for $1-5$~ min at a pressure of $1-5$~ mbar to form the tunnel barrier of the NIS probe. Next, approximately 25 nm Cu at $\theta=-14^\circ$ is evaporated to complete the NIS junction and to form the normal metal wire. Figure~\ref{fig:SEM}\figta shows a scanning electron microscope (SEM) image of a typical Nb-SQUIPT device with an enlarged view of the zone around the weak link. The measurements of the samples have been performed in a $^3$He/$^4$He dilution refrigerator~\cite{Pekola1994} down to the base temperature of about 80 mK. 
\section{Results and discussion}
The relevant dimensions of a typical measured SNS structure [see inset of Fig.~\ref{fig:SEM}\figb] are the minimum Nb electrode separation $L_1=0.49~ \mu \mr m$,  the full length $L_2=2.15~ \mu \mr m$ and  minimum width $w=0.54~ \mu \mr m$ of the Cu wire.  We measure typical low temperature resistance value of the Cu strip to be about $0.4~\mr\Omega$. In the diffusive SNS junction, the Thouless energy $E_\mr{Th}=\hbar D/L^2$ determines the energy scale for the proximity effect. Here $D$ and $L$ are the diffusion constant and the effective length of the Cu wire, respectively~\cite{Giazotto2008,Blum2004,Dubos2008}. The effective length of the normal metal $L=1.2~\mr{\mu m}$ is derived from the estimated values of $E_\mr{Th}\simeq 4.5 ~\mr{\mu eV}$ and the measured Cu diffusion constant $D=0.01~\mr{m^2s^{-1}}$~\cite{Najafi2016}. The Nb-Cu-Nb sample considered here  is in the long junction limit $L\approx16~ \xi_0$, where $\xi_0=\sqrt{\hbar D/\mr\Delta}$ is the coherence length and $\mr\Delta\approx1.2~\mr{meV}$  the superconducting energy gap of Nb. 

The current-voltage characteristic of the Nb-Cu-Nb structure is shown in the main panel of Fig.~\ref{fig:SEM}\figtb at different values of bath temperatures. The dc  voltage $V_\mr{SNS}$ is measured across the structure biased by a  current $I_\mr{SNS}$ in a four probe configuration as indicated in the inset of Fig.~\ref{fig:SEM}\figb. 
As shown by the IV curves, the hysteretic behavior from self-heating in the finite-voltage state is observable at $T<650~\mr{mK}$ and hence the magnitudes of switching $I_\mr{sw}$ and retrapping $I_\mr{r}$ current differ from each other. At the base temperature, we find them to be about $30~\mr{\mu A}$ and $6~\mr{\mu A}$, respectively. 


An IV characteristic of the Nb-SQUIPT is illustrated in Fig.~\ref{fig:IV}\figa,  measured at two different magnetic fields, $\mr\Phi=0$ (red) and $\mr\Phi=0.5 ~\Phi_0$ (blue), corresponding to maximum and minimum  minigap opened in the normal weak link~\cite{Sueur2008,Zhou1998}. Furthermore, the flux-dependent onset of current on the quasiparticle branch is shown in the inset  when $|V|$  exceeds the sum of the probe electrode gap $\mr\Delta_\mr{Al}/e$ and the minigap induced in the N part.
In the diffusive regime of a metallic SNS junction, the minigap in the normal metal is of the order of the Thouless energy $E_\mr{Th}$~\cite{Pannetier2000}.
Considering the bias voltage dependence of the differential conductance~\cite{Meschke2011,Najafi2014} 
and subtracting the contribution of an effective series resistance $R_\mr S=1$  k${\mr\Omega}$, arising from the two point measurement technique,
the magnitudes of the superconducting probe electrode gap and minigap are  $\mr\Delta_\mr{Al}=253~~\mu\mr{eV}$ and $\mr\Delta_\mr{g}=27~\mu\mr{eV}$, respectively.
As a consequence of the minigap variations, the current  $I (\mr\Phi)$ is periodic as a function of the flux through the loop. Typical current modulations are displayed in Fig.~\ref{fig:IV}\figtc for some selected values of bias voltages $V$ in the quasiparticle branch from $250~\mu \mr V$ to $320~\mu \mr V$. Figure~\ref{fig:IV}\figtc indicates that the measured current modulation reaches a peak-to-peak amplitude as large as $\delta I=5.5~ \mr {nA}$ at the bias voltage $V=300~\mu \mr V$. 
We now consider the flux modulation in the low bias regime. Figure~\ref{fig:subgap}\figta \- demonstrates the extent of the flux modulation of the IV curve in the sub-gap regime close to zero bias at two different temperatures, $T=80$ mK (blue) and $T=190$ mK (red). At $T=190$ mK, almost full magnetic flux modulation of the IV is visible due to increased SNS weak link inductance in comparison to the loop inductance~\cite{Ronzani2014}. 
We can furthermore observe current peak at the bias voltage  $V=30$ $\mu$V  
with the maximum current $I_{\max}=3.6$ nA. It is caused by multiple Andreev reflections, which are visible due to
relatively low tunnel resistance of the probe junction, $R_{\mr T}=1$ k${\mr \Omega}$.
 In this sub-gap regime, the current modulation as a function of magnetic flux $I (\mr\Phi)$ is shown in Fig.~\ref{fig:subgap}\figtb at few selected values of applied bias voltage.  At $|V|\approx 30~\mu \mr V$, we find the maximum peak-to-peak amplitude as large as $\delta I\approx 2.5~ \mr{nA}$,  corresponding to the maximum sensitivity $|\partial I/\partial \mr\Phi|_\mr{max}\approx 55~ \mr{nA/\Phi_0}$. The effect of temperature $T$ on the flux-to-current transfer function is displayed in Fig.~\ref{fig:subgap}\figc, showing the temperature dependence of the maximum sensitivity in the low bias regime. It decreases monotonously with increasing $T$. For comparison, the transfer functions close to the optimum bias values in the supercurrent and  quasiparticle regions are plotted in Fig.~\ref{fig:res_flux}.

 We now discuss the bias voltage dependence of the maximum sensitivity $|\partial I/\partial\mr\Phi|_\mr{max}$ at the base temperature $T\approx 80~ \mr{mK}$. Figure~\ref{fig:res_vbias} indicates that the maximum current responsivity is obtained at $|V|\simeq30~\mu \mr V$, corresponding to $|\partial I/\partial\mr\Phi|_\mr{max}\approx 55~\mr {nA/\Phi_0}$. At this working point, the typical output current level is $I\approx 3~\mr{nA}$ implying an average total power dissipation for the Nb-SQUIPT to be below $\mr{10^{-13}~W}$. The value can be further reduced by simply increasing the resistance of the probing junction. This power is two orders of magnitude smaller than in our previous device with an Al superconducting probe~\cite{Najafi2014}.  The sub-gap operation of a SQUIPT is thus a possible choice for applications where very low dissipation is required.
\begin{figure}
  \includegraphics[width=1\textwidth]{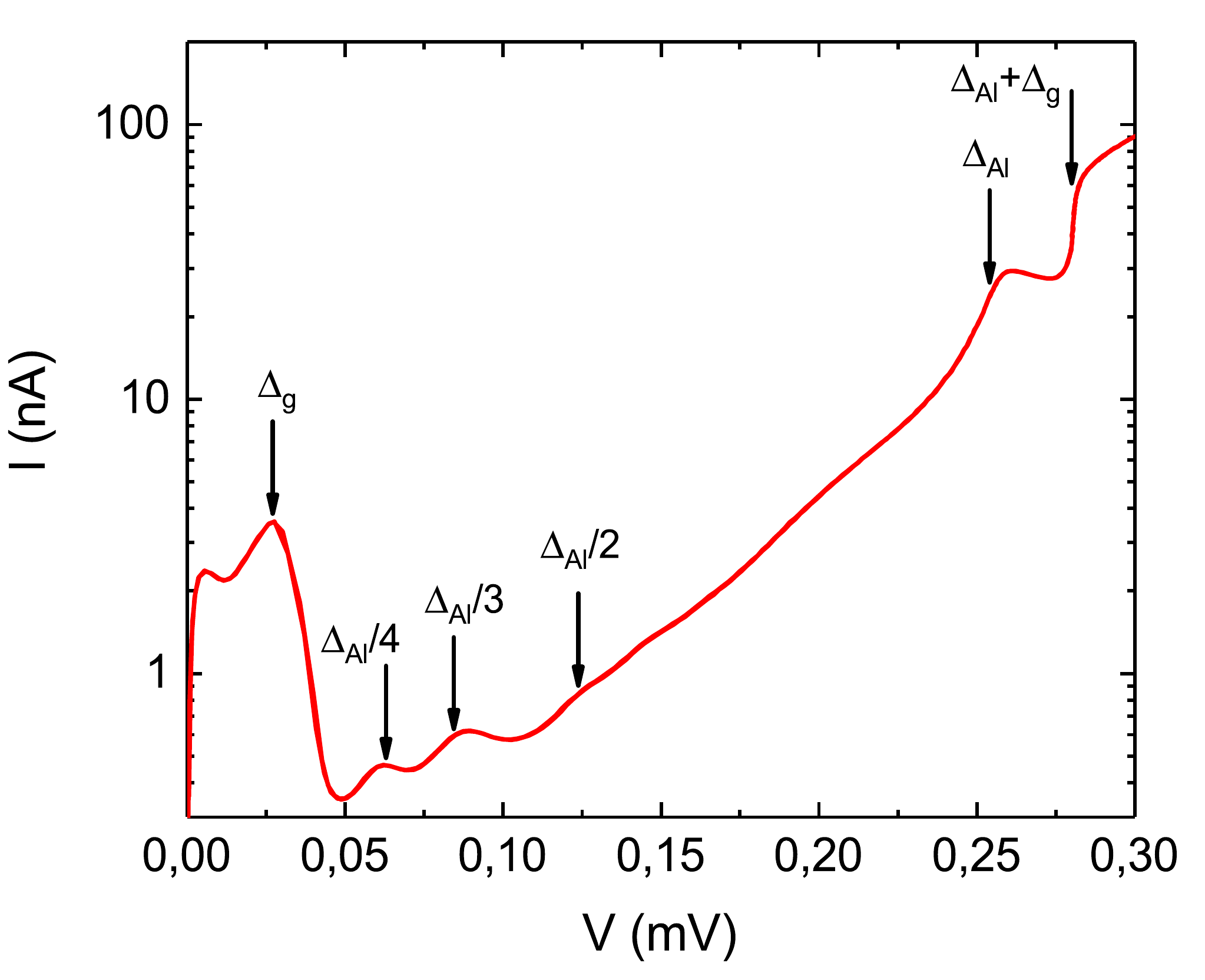}
\centering
\caption{Subgap structure related to multiple Andreev reflections in the I-V curve taken at zero magnetic field and at $T=78$ mK. 
The contribution of a 1 k${\mr\Omega}$ series resistor has been subtracted. } 
\label{fig:res_MAR}       
\end{figure}
Let us now briefly discuss the mechanism behind the SQUIPT response to the magnetic field at low bias.
Considering first an ideal case, we assume that the response comes from the Josepshson current of the probe
junction, which is modulated by the field together with the minigap in the copper island. 
The expected value of the critical current may be estimated by means of the Ambegaokar-Baratoff formula \cite{AB}
for an asymmetric junction with the gaps of the leads ${\mr \Delta}_{\rm g}({\mr \Phi})$ and ${\mr \Delta}_{\rm Al}$.
With the junction resistance $R_\mr T=1$ k${\mr\Omega}$ we find a maximum critical current
$I_\mr C=150$ nA.

 Assuming now that the minigap is modulated as 
${\mr \Delta}_{\mr g}({\mr \Phi}) \approx {\mr \Delta}_{\mr g}\left|\cos(\pi{\mr \Phi}/{\mr \Phi}_0)\right|$,
we can estimate the maximum responsivity of the device as $\sim\pi I_\mr C/{\mr\Phi}_0\approx 500$ nA/${\mr\Phi}_0$,
which is about 10 times higher than in the experiment. Unfortunately, in our sample we observe only a very small 
critical current $I_\mr C= 2.4$ nA, which, on top of that, is not modulated by flux, and this simple mechanism is not working.
However, our analysis shows that improving the quality of the probe junction may potentially 
increase the sensitivity of the SQUIPT. 
We believe that the observed low-bias current modulation in our device comes from multiple Andreev reflections (MAR). 
The tunnel barrier of the probe junction is rather transparent (overlap size $\mr{310~nm\times 510~ nm}$,  characteristic resistance $\mr{\approx150~\Omega \mu m^2}$) and MAR signatures are clearly visible 
in the I-V curve plotted in Fig. \ref{fig:res_MAR}. The observed pattern  agrees well with the one expected
for a junction between two different superconductors with the gaps ${\mr\Delta}_{\rm g}$ and ${\mr\Delta}_{\rm Al}$ \cite{Cuevas}.
Namely, we find current peaks at $eV={\mr\Delta}_{\rm g}$, $eV={\mr\Delta}_{\rm Al}$, a current jump at $eV={\mr\Delta}_{\rm Al}+{\mr\Delta}_{\rm g}$,
and a series of less visible features at voltages ${\mr\Delta}_{\rm Al}/2e$, ${\mr\Delta}_{\rm Al}/3e$ etc. 
In terms of responsivity at low bias, the MAR peak centered around ${\mr\Delta}_{\rm g}/e=27$ ${\rm \mu}$V is important. 
The position of this peak follows the magnetic field dependent minigap ${\mr\Delta}_{\rm g}({\mr\Phi})$, which results
in the modulation of the current at fixed voltage shown in Fig. \ref{fig:subgap}(b).
The peak is smeared by increasing temperature and eventually vanishes at $T>{\mr\Delta}_{\rm g}/k_\mr B\approx 300$ mK,
see  Fig. \ref{fig:subgap}(c).

\section{Summary}
To summarize, we have investigated Nb-Cu-Nb weak links and, consequently, a Nb-SQUIPT at low bias voltages of the tunnel probe. The structures are based on a fabrication process with two independent lithography and deposition steps, relying on Ar ion cleaning of the Nb contact surfaces. The typical power dissipation in the fW range gives the opportunity of using the Nb-SQUIPT as a low dissipation magnetometer and, furthermore,  in other detector applications such as ultrasensitive bolometers and calorimeters.\\

\textbf{Acknowledgments}\\\

The work has been supported by the Academy of Finland (project numbers 284594 and 275167). We acknowledge Micronova Nanofabrication Centre of OtaNano research infrastructure for providing the processing facilities, and  for the sputtered Nb films.

\end{document}